\documentclass[]{aa}
\usepackage{graphicx}
\usepackage{lscape}
\usepackage{natbib}
\usepackage{hyperref}
\usepackage{bigstrut} 
\usepackage{amsmath}
\usepackage[varg]{txfonts}
\begin{document}

\newcommand{\Lsun}{$L_{\odot}$}
\newcommand{\Msun}{$M_{\odot}$}

\title{Large-scale environments of binary AGB stars probed by \emph{Herschel}\thanks{{\it Herschel} is an ESA space observatory with science instruments provided by European-led Principal Investigator consortia and with important participation from NASA. This paper makes use of data from ESO programmes 076.D-0624, 077.D-0620, 078.D-0122, 080.D-0076, 187.D-0924.}}
\subtitle{II. Two companions interacting with the wind of $\pi^1$~Gruis}

   \author{A.~Mayer
          \inst{1}
          \and A.~Jorissen
          \inst{2}
          \and C.~Paladini
          \inst{2} 
          \and F.~Kerschbaum
          \inst{1}
          \and D.~Pourbaix
          \inst{2}\fnmsep\thanks{Senior Research Associate, F.R.S.-FNRS, Belgium} 
          \and C. Siopis
          \inst{2}
          \and R.~Ottensamer
          \inst{1}
          \and M.~Me\v{c}ina
          \inst{1}         
          \and N.L.J.~Cox
          \inst{3} 
          \and M.A.T.~Groenewegen
          \inst{4}
          \and D.~Klotz
          \inst{1}
          \and G. Sadowski
          \inst{2}
          \and A. Spang
          \inst{5}
          \and P. Cruzal{\`e}bes
          \inst{5}
          \and  C.~Waelkens
          \inst{3}
          }

\institute{University of Vienna, Department of Astrophysics, Sternwartestra\ss e 77, 1180 Wien, Austria \\
              \email{a.mayer@univie.ac.at}
\and
Institut d'Astronomie et d'Astrophysique, Universit\'e Libre de Bruxelles, CP. 226, Boulevard du Triomphe, 1050 Brussels, Belgium
\and
Instituut voor Sterrenkunde, KU Leuven, Celestijnenlaan, 200D, 3001 Leuven, Belgium  
\and
Koninklijke Sterrenwacht van Belgi\"e, Ringlaan 3, 1180 Brussels, Belgium
\and
Laboratoire Lagrange, UMR 7293, Universit\'e de Nice-Sophia Antipolis, CNRS, Observatoire de la C\^ote d’Azur, Bd de l’Observatoire, CS 34229,
F-06304 Nice cedex 4, France
 }

   \date{Received 24 June 2014; accepted 14 August 2014}

 \abstract 
{The \textit{\textup{Mass loss of Evolved StarS}} (MESS) sample observed with PACS on board the Herschel Space Observatory revealed that several asymptotic giant branch (AGB) stars are surrounded by an asymmetric circumstellar envelope (CSE) whose morphology is most likely caused by the interaction with a stellar companion. The evolution of AGB stars in binary systems plays a crucial role in understanding the formation of asymmetries in planetary nebul\ae~(PNe), but at present, only a handful of cases are known where the interaction of a companion with the stellar AGB wind is observed.}
{We probe the environment of the very evolved AGB star $\pi^1$~Gruis on large and small scales to identify the triggers of the observed asymmetries.}
{Observations made with \emph{Herschel}/PACS at 70\,$\mu$m and 160\,$\mu$m picture the large-scale environment of $\pi^1$~Gru. The close surroundings of the star are probed by interferometric observations from the VLTI/AMBER archive. An analysis of the proper motion data of Hipparcos and Tycho-2 together with the Hipparcos Intermediate Astrometric Data help identify the possible cause for the observed asymmetry.}
{The \emph{Herschel}/PACS images of $\pi^1$~Gru show an elliptical CSE whose properties agree with those derived from a CO map published in the literature. In addition, an arc east of the star is visible at a distance of $38\arcsec$ from the primary. This arc is most likely part of an Archimedean spiral caused by an already known G0V companion that is orbiting the primary at a projected distance of 460\,au with a period of more than 6200\,yr. However, the presence of the elliptical CSE, proper motion variations, and geometric modelling of the VLTI/AMBER observations point towards a third component in the system, with an orbital period shorter than 10\,yr, orbiting much closer to the primary than the G0V star.}
{}

\titlerunning{Large-scale environments of binary AGB stars probed by \emph{Herschel} -- II.}
\maketitle
%

\section{Introduction}

The evolution of low- and intermediate-mass stars ends with an ascent of the asymptotic giant branch (AGB) in the Hertzsprung-Russell diagram. This phase involves an increase of mass loss that strips off the envelope through a slow and dust-enriched wind ($v_{\rm w}=5-20$\,km\,s$^{-1}$) blown into the interstellar medium (ISM), where it ranks among the dominant contributors of heavy elements in the Galaxy. Finally, at the end of the AGB phase, the remaining envelope as a whole is ejected. The hot remnant stellar core ionizes the ejecta, forming what is known as a planetary nebula (PN). PNe show a manifold of morphological diversity, including highly asymmetric and bipolar forms that can only be adequately described by a binary star model \citep[e.g.][]{Nordhaus2007,DeMarco2008,Miszalski2009a,Miszalski2009b}.

\citet{Paczynski1971}, \citet{Livio1988}, and \citet{Theuns1993} have theoretically shown that already in the AGB phase the stellar winds must be heavily distorted in binary systems depending on the size of the system and the evolutionary type of the companion. For binary systems with small separations, the primary AGB star fills the Roche lobe and transfers mass onto the companion \citep{Paczynski1965,Paczynski1971}. 

In detached systems, the stellar AGB wind fills the Roche lobe, and up to half of the material is accreted by the companion \citep[named \textit{\textup{wind Roche lobe overflow}  -- WRLOF},][]{Mohamed2011,Abate2013}. Jets and bipolar outflows have also been observed for some of these systems, e.g., \object{$o$~Cet} \citep{Meaburn2009}, \object{R~Aqr} \citep{Wallerstein1980,Kafatos1989}, and \object{V~Hya} \citep{Hirano2004}, where the accretion disc of the companion is fed by a strong wind from the AGB primary \citep[e.g.][]{Morris1987,Soker2000,Huggins2007}. But in general, the companion affects the circumstellar envelopes of the AGB star in two ways. First, the material that is transferred via the WRLOF is focused by the gravitational potential of the companion and forms a density wake that trails the orbital motion of the companion. The result is a wind pattern shaped as an Archimedean spiral, as predicted by hydrodynamic simulations in \citet{Theuns1993}, \citet{Mastrodemos1998,Mastrodemos1999}, and \citet{Kim2012a}. Second, the presence of the companion also manifests itself via the gravitational force that it exerts on the primary, causing it to move around the centre of mass of the binary system \citep{Soker1994,Kim2012b,Kim2012c}. Furthermore, \citet{Kim2012b}  and \citet{Kim2013} recently demonstrated that the combination of the two effects leads to a spiral wind pattern exhibiting knots where the two structures intersect.

Observationally, spiral patterns were found around a small number of AGB or proto-PN objects: \object{AFGL~3068} \citep{Mauron2006}, \object{CIT~6} \citep{Dinh2009,Kim2013}, $o$~Cet \citep{Mayer2011}, \object{R~Scl} \citep{Maercker2012}, and \object{W~Aql} \citep{Mayer2013}, all of which are wide binary systems with an orbital separation in the range of $\approx$50--160\,au. Recently, \citet{Mauron2013} found that 50\% of a sample of 22 AGB stars have elliptical emission, which the authors attributed to binaries whose envelopes are flattened by a companion.

This work continues our study of large-scale environments of binary AGB stars from the \emph{Herschel} Mass loss of Evolved StarS sample \citep[MESS;][]{Groenewegen2011}. Contrary to \citet[][hereafter Paper~I]{Mayer2013}, which concentrated exclusively on \emph{Hersche}/PACS observations of the large-scale structures (around R~Aqr and W~Aql), here we also explore the close surroundings of the star using Hipparcos Intermediate Astrometry Data \citep[IAD;][]{vanLeeuwen1998} as well as unpublished archive observations obtained with the Very Large Telescope Interferometer Astronomical Multi-BEam combineR \citep[VLTI/AMBER,][]{petrov2007}. We do this by analysing the structures around the binary AGB star \object{$\pi^1$~Gru} at angular scales from $0\farcs02$ to $60\arcsec$. Section~\ref{Sect:properties} discusses the fundamental properties of $\pi^1$~Gru. In Section~\ref{Sect:observations}, the observations of \emph{Herschel}/PACS and VLTI/AMBER are described, with their results presented in Section~\ref{Sect:results}. The different interaction scenarios that can produce asymmetries in the extended environment of the star on both small and large scales are discussed in Section~\ref{Sect:discussion}.

\section{General properties of $\pi^1$~Gruis}
\label{Sect:properties}

\begin{figure}[t!]
\centering
\includegraphics[width=9cm]{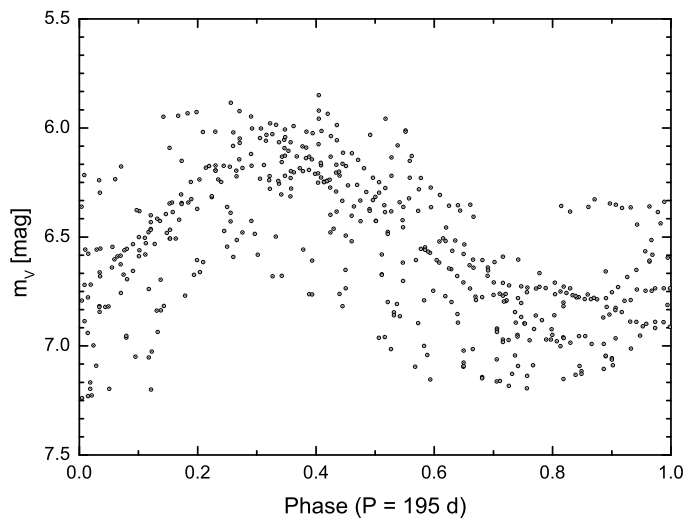} 
\caption{ASAS-3 light curve of $\pi^1$~Gru with an adopted pulsation period of 195 days covering 14 cycles.}
\label{light_curve}     
\end{figure}

The S5,7 star $\pi^1$~Gru (HIP~110478) is an SRb variable, and because of its proximity \citep[$d = 163$~pc;][]{vanLeeuwen2007} one of the brightest and best studied intrinsic\footnote{About intrinsic S stars, see \citet{VanEck1999}.} S~stars \citep{Keenan1954}. The intrinsic nature of the S~star $\pi^1$~Gru is defined from the presence of spectral lines of the element Tc \citep{Jorissen1993}. The pulsation period of the star was initially derived to be $\approx$150 days by \citet{Eggen1975} and used in various publications since then. A new derivation of the period based on the light curve provided by the All Sky Automated Survey (ASAS) Photometric V-band Catalogue \citep{Pojmanski2005} revealed , however, $\pi^1$~Gru varies with a pulsation period of 195 days (see Fig.\ref{light_curve}).

\begin{figure}[t!]
\centering
\includegraphics[width=9cm]{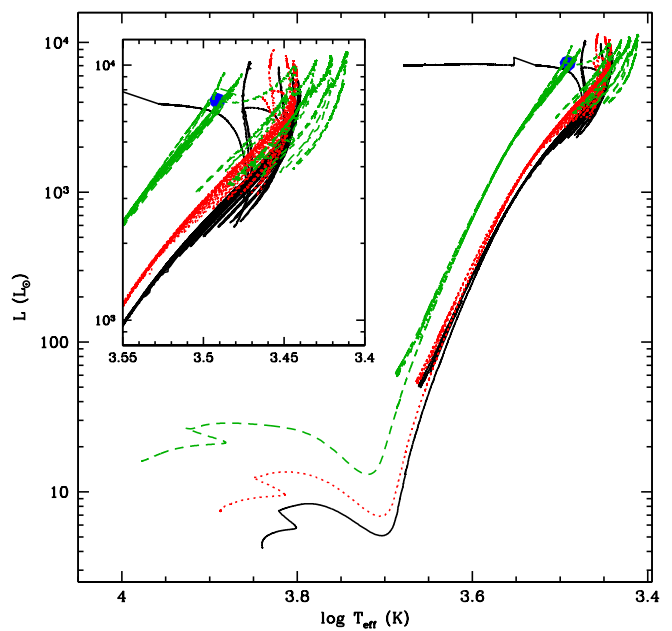} 
\caption{Evolutionary tracks from STAREVOL for stars with initial masses 1.5\,$M_\sun$ (black line), 1.7\,$M_\sun$ (red dotted line) and 2.0\,$M_\sun$ (green dashed line) from the pre-main-sequence to the end of the AGB. $\pi^1$~Gru is represented by the large blue circle.
\label{Fig:HRD}}     
\end{figure}

The Hertzsprung-Russell diagram for Hipparcos S~stars from \citet{vanEck1998} point to the very evolved nature of this star, close to the tip of the AGB. With values of $T_{\rm eff}=3100$\,K and $\log L/L_\sun = 3.86$ \citep{vanEck1998}, the location of $\pi^1$~Gru in the Hertzsprung-Russell diagram can be compared with evolutionary tracks (Fig.~\ref{Fig:HRD}) computed from the STAREVOL code \citep{Siess2006,Siess2008} with a metallicity $Z=0.02$. It appears that $\pi^1$~Gru falls on the track of a star of initial mass 2.0\,$M_\sun$, but by the time that star has reached $\log L/L_\sun = 3.86$, its mass has dropped to about 1.5\,$M_\sun$.

From CO(1--0) observations, a present-day mass loss rate of $\dot{M}=2.73\times10^{-6}$\,$M_\sun$\,yr$^{-1}$ \citep{Winters2003}, an expansion velocity of 14.5\,km\,s$^{-1}$ \citep{Guandalini2008}, and a gas-to-dust ratio of 380 \citep{Groenewegen1998} is derived.
\begin{table*}[t!]
\caption{\label{tab:amber_obs}Journal of the AMBER observations of $\pi^{1}$~Gru.}
\begin{center}
\begin{tabular}{llllll}
\hline
\hline
Date and UT time & Config. & Baselines & PA & Seeing & Airmass \\ 
 & & [m]  & [deg]  & [$\arcsec$] &  \\
\hline
    09 Oct 2007 T23:51:44.10&E0-G0-H0   &16 - 31 - 47&230&0.68-0.71&1.18\\
    10 Oct 2007 T00:53:41.50&   "       &16 - 32 - 48   &242&0.70-0.78&1.09\\
    10 Oct 2007 T01:05:39.23&   "       &16 - 32 - 48   &244&0.67-0.63&1.08\\
    10 Oct 2007 T01:50:37.99&   ''      &16 - 32 - 48   &252&1.05-1.10&1.07\\
    10 Oct 2007 T03:05:37.54&  ''       & 15 - 30 - 45   &265& 0.67-0.68&1.11\\
    10 Oct 2007 T03:54:47.47&   "       &14 - 28 - 42   &274&0.56-0.47&1.18\\
    10 Oct 2007 T04:39:09.68&   "       &13 - 26 - 39   &283&0.57-0.56&1.30\\
\hline
\end{tabular}
\end{center}
\end{table*}

$\pi^1$~Gru is known to have a faint G0V companion with an apparent visual magnitude of 10.4 \citep{Feast1953,Ake1992}. From the Hipparcos parallax \citep[$6.13\pm0.76$~mas or 163\,pc;][]{vanLeeuwen2007}, the distance modulus is 6.1, yielding an absolute visual magnitude of 4.3 for the G0V star, in accordance with its spectral classification. The companion must thus be physically associated with the S~star, but the orbital period is quite long, since the relative position did not change significantly over the past century, according to the list of relative positions collected by the Washington Double Star Observations catalogue, and kindly communicated to us by~B.~Mason (see Appendix~\ref{App:Pos.G0V} and Table~\ref{Tab:pi_gru_pos}). Assuming that the observed angular separation ($\approx 2\farcs8$) corresponds to the semi-major axis and adopting 2.5\,$M_{\sun}$ as the total mass of the system ($\pi^1$~Gru+G0V companion), the system parallax implies an orbital separation of the order of 460\,au and an orbital period of about 6200\,yr.

\section{Observations}
\label{Sect:observations}

\subsection{Herschel/PACS}

The observations presented here are part of the MESS Guaranteed Time Key Programme \citep{Groenewegen2011} for the \emph{Herschel} Space Observatory \citep{Pilbratt2010} using the Photodetector Array Camera and Spectrometer \citep[PACS,][]{Poglitsch2010} and the Spectral and Photometric Image Receiver \citep[SPIRE,][]{Griffin2010} on board the spacecraft. $\pi^1$~Gru was observed on May 21 2010. For the following analysis, we exclusively used the PACS data since the instrument offers a resolution that best suits our purpose. Observations of $\pi^1$~Gru were obtained at 70 and 160\,$\mu$m at a FWHM of $5\farcs6$ and $12\arcsec$, respectively. The adopted data processing and image reconstruction for $\pi^1$~Gru  was made in the same way as for the data presented in Paper~I, following  \citet{Groenewegen2011} and \citet{Roussel2013}. We oversampled the reconstructed images by a factor 3.2 to achieve a pixel size of $1\arcsec$ in the blue and $2\arcsec$ in the red band. 

An overview of the MESS objects is given by \citet{Cox2012} and detailed studies of individual objects are presented in \citet{Ladjal2010}, \citet{Kerschbaum2010}, \citet{Jorissen2011}, \citet{Mayer2011,Mayer2013}, \citet{Decin2011,Decin2012}, \citet{vanHoof2013}, and \citet{Mecina2014a}.

\subsection{VLTI/AMBER}
\label{Sect:Amber_Obs}

$\pi^{1}$~Gru was observed with VLTI/AMBER and VLTI/MIDI in the framework of ESO programmes 076.D-0624, 077.D-0620, 078.D-0122, 080.D-0076, 
and 187.D-0924. The detailed description of the MIDI data reduction and modelling is given in \citet{Sacuto2008} and Paladini et al. (in prep).
The MIDI observations do not deviate from spherical symmetry. These data, however, sample only the low spatial frequencies,
and it is known that for AGBs, asymmetric structures are usually detected at high spatial frequencies. 
The only information that we can extract from the MIDI observations is the overall size of the envelope. For this reason the MIDI data are not discussed here.
\begin{figure}
\begin{center}
\includegraphics[width=0.5\textwidth, bb = 105 382 549 711]{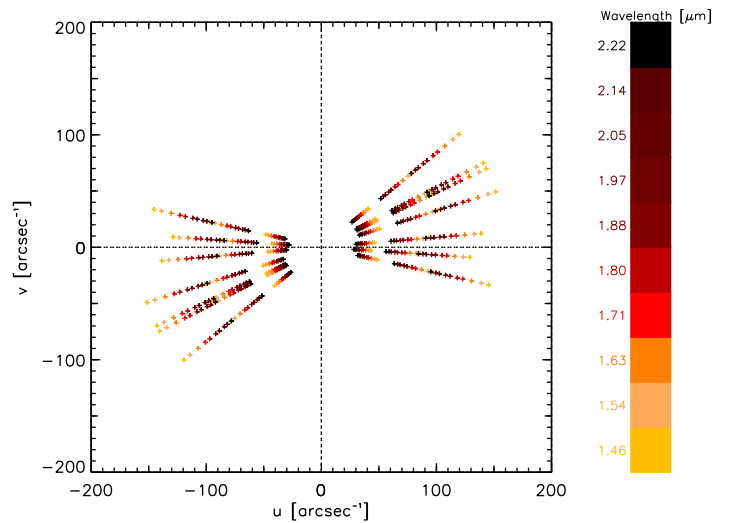}
\caption{$uv$-plane coverage of the VLTI/AMBER data of $\pi^1$~Gru. }
\label{fig:uv}
\end{center}
\end{figure}

We retrieved seven VLTI/AMBER observations from the ESO archive. The data were recorded in low-resolution mode ($R=30$) on the night of October 9, 2007 and cover the $J$, $H$, and $K$ bands. The log of these observations is presented in Table~\ref{tab:amber_obs}, while Fig.~\ref{fig:uv} shows the corresponding $uv$-plane coverage. 

We reduced the data with amdlib v.3.0.8 \citep{Tatulli2007,Chelli2009} using the K3III star \object{$\lambda$~Gru} as calibrator \citep[with a stellar diameter of $2.62\pm0.03$\,mas]{Borde2002,Cruzalebes2010}. The data analysis is limited to the $H$ and $K$ bands because reliability for the wavelengths shorter than 1.46\,$\mu$m is not guaranteed by the current pipeline version. 

The medium- and long-baseline visibilities sample the second and even third lobe, meaning that the star is fully resolved. There is evidence for deviation from centro-symmetry (or rather left/right symmetry for this triplet of aligned baselines), as judged from the non-zero closure phase. We return to this in Sect. 5.3.2. 

\section{Results}
\label{Sect:results}

\begin{figure*}[t!]
\centering
   \includegraphics[width=9cm]{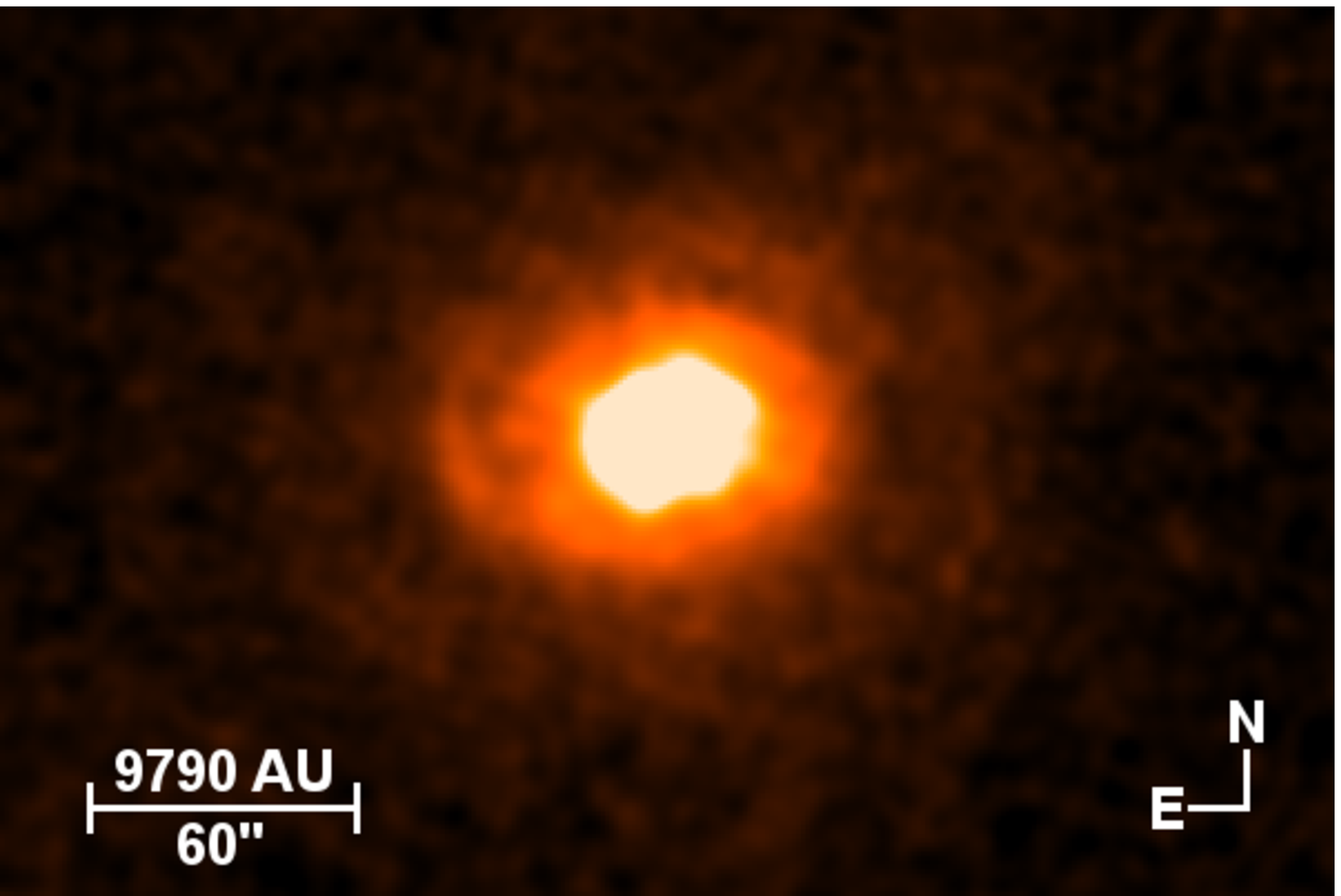} 
   \includegraphics[width=9cm]{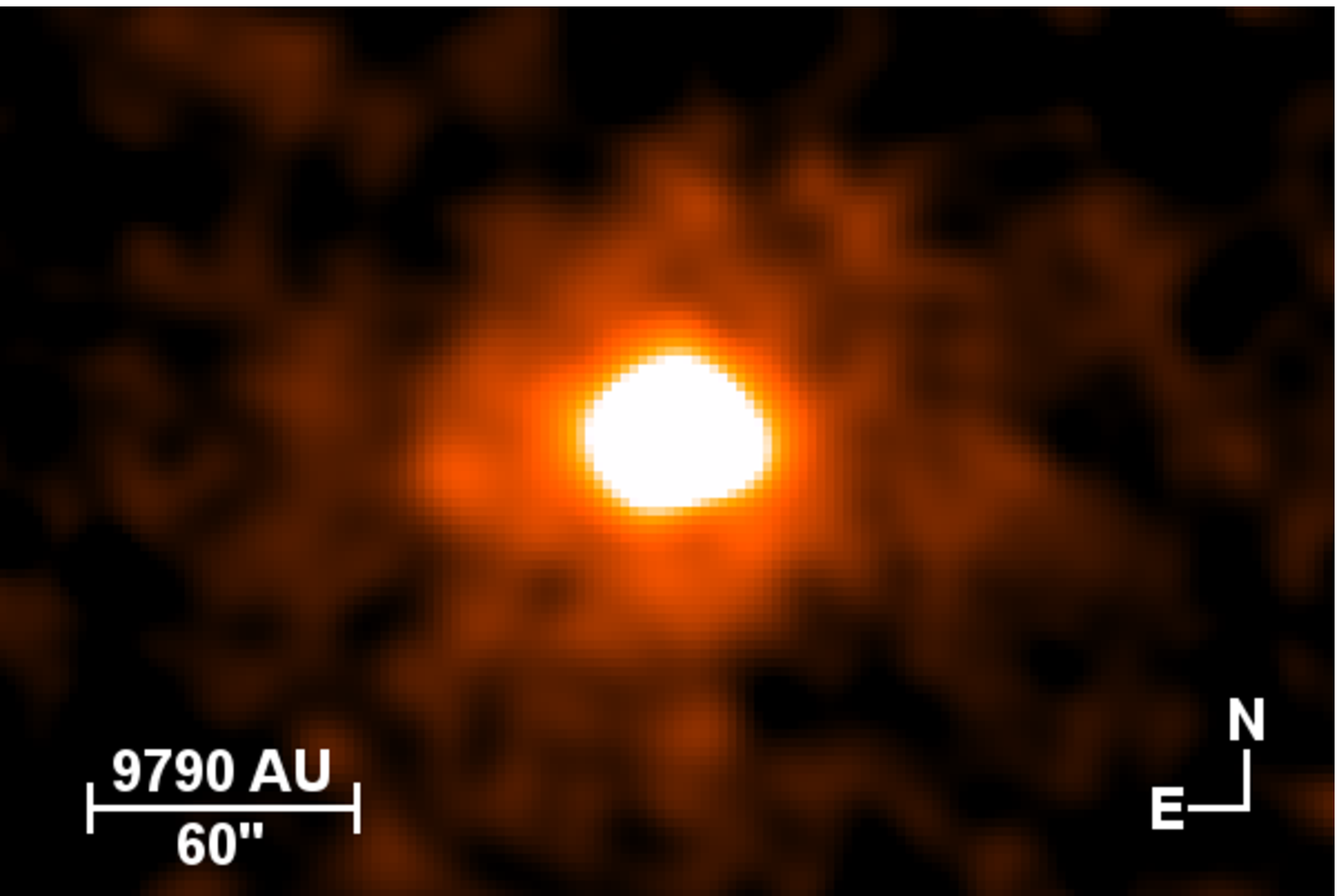}
     \caption{Deconvolved \emph{Herschel}/PACS images of $\pi^1$~Gru at 70\,$\mu$m (left panel) and 160\,$\mu$m (right panel).}
\label{pi_gru_pacs}
\end{figure*}

Figure~\ref{pi_gru_pacs} depicts the images obtained with \emph{Herschel}/PACS at 70\,$\mu$m (blue band) and 160\,$\mu$m (red band)\footnote{Both maps are available as FITS files from CDS/VizieR at \url{http://cdsweb.u-strasbg.fr/}}. Since the blue band offers a better spatial resolution, all following discussions are based on this image unless stated otherwise. The PACS 70\,$\mu$m image is dominated by two features, an elliptical CSE and an arc east of the star, which are described in the remainder of this section. 

The elliptical CSE is oriented east-west with its major axis at PA~$\approx105\degr$. The total size of the emission is $\approx 72\arcsec \times\; 60\arcsec$ [$11750\times9790$\,au]. A confirmation of the CSE size is provided by the 870\,$\mu$m image of $\pi^1$~Gru obtained with the APEX bolometer LABoCa \citep{Ladjal2010b}. The authors found an elongation of the CSE in east-west direction with a total size of the structure of about $60\arcsec \times 40\arcsec$.  

The size of the CSE obtained from the PACS 70\,$\mu$m image is much smaller than that inferred by \citet{Young1993b} from the IRAS 60\,$\mu$m dust emission ($4\farcm9$). A similar discrepancy for the CSE sizes obtained from IRAS and \emph{Herschel} data was found for the targets X~Her and TX~Psc analysed previously \citep{Jorissen2011}, most likely owing to the size of the IRAS PSF (FWHM $=1\farcm6$ at $60\,\mu$m), which causes difficulties in probing asymmetries of the order of $1\arcmin$ as found in the \emph{Herschel} data. 

The second main feature visible in the far-IR emission of $\pi^1$~Gru is an arc east of the central star. It emerges $38\arcsec$ away from the stellar system, in the direction of the major axis ($\approx 105\degr$). This finding diminishes the probability that the elliptical emission is caused by the interaction with the ISM. The observed arc is curved towards the north and extends in that direction for almost $25\arcsec$ and $16\arcsec$ to the east. On the PACS 160\,$\mu$m image, the arc suffers from the low resolution but remains recognisable as a clump. We further note that in Fig.~A.2 of \citet{Ladjal2010b}, a spike is visible on the eastern side, that might reflect the arc seen on the $70\,\mu$m PACS image.

\section{Discussion}
\label{Sect:discussion}

\subsection{Origin of the far-infrared dust arc}
\label{Sec:arc}

Arcs or arms around AGB stars were found recently for CIT~6 \citep{Dinh2009,Kim2013}, \object{TX~Cam} \citep{Castro2010}, $o$~Cet \citep{Mayer2011}, and W~Aql \citep{Mayer2013}. They are most likely part of Archimedean spirals caused by a combination of (i) the accretion wake of the companion when it orbits the mass-losing primary and the wind pushes the wake radially outwards \citep{Theuns1993,Mastrodemos1998,Mastrodemos1999} and (ii) the motion of the primary around the centre of mass caused by the gravitational drag of the companion \citep{Soker1994,Kim2012b,Kim2012c}. Spectacular examples of complete  Archimedean spirals have been observed around AFGL~3068 \citep{Mauron2006} and R~Scl \citep{Maercker2012}.

It is not clear, however, whether the far-infrared arc seen around $\pi^1$~Gru has anything to do with a spiral pattern produced by the orbital motion of a companion. If it does, the properties of the arc have to be consistent with the properties of the $\pi^1$~Gru+G0V binary system. From the orbital period $P\approx6200$\,yr, the parallax $\varpi=6.13$\,mas, and the wind velocity $v_{\rm w}=14.5$\,km\,s$^{-1}$ , one derives an arm separation of
\begin{equation}
\rho = v_{\rm w} P \varpi = 116\arcsec.
\label{arm_separation}
\end{equation}
This is more than twice as large as the separation of $52\arcsec$ between the central part of the arc and $\pi^1$~Gru as seen on the PACS 70\,$\mu$m image. If the arc is part of a spiral, it only represents a part of the first spiral twist. 

\begin{figure}[t!]
\centering
   \includegraphics[width=9cm]{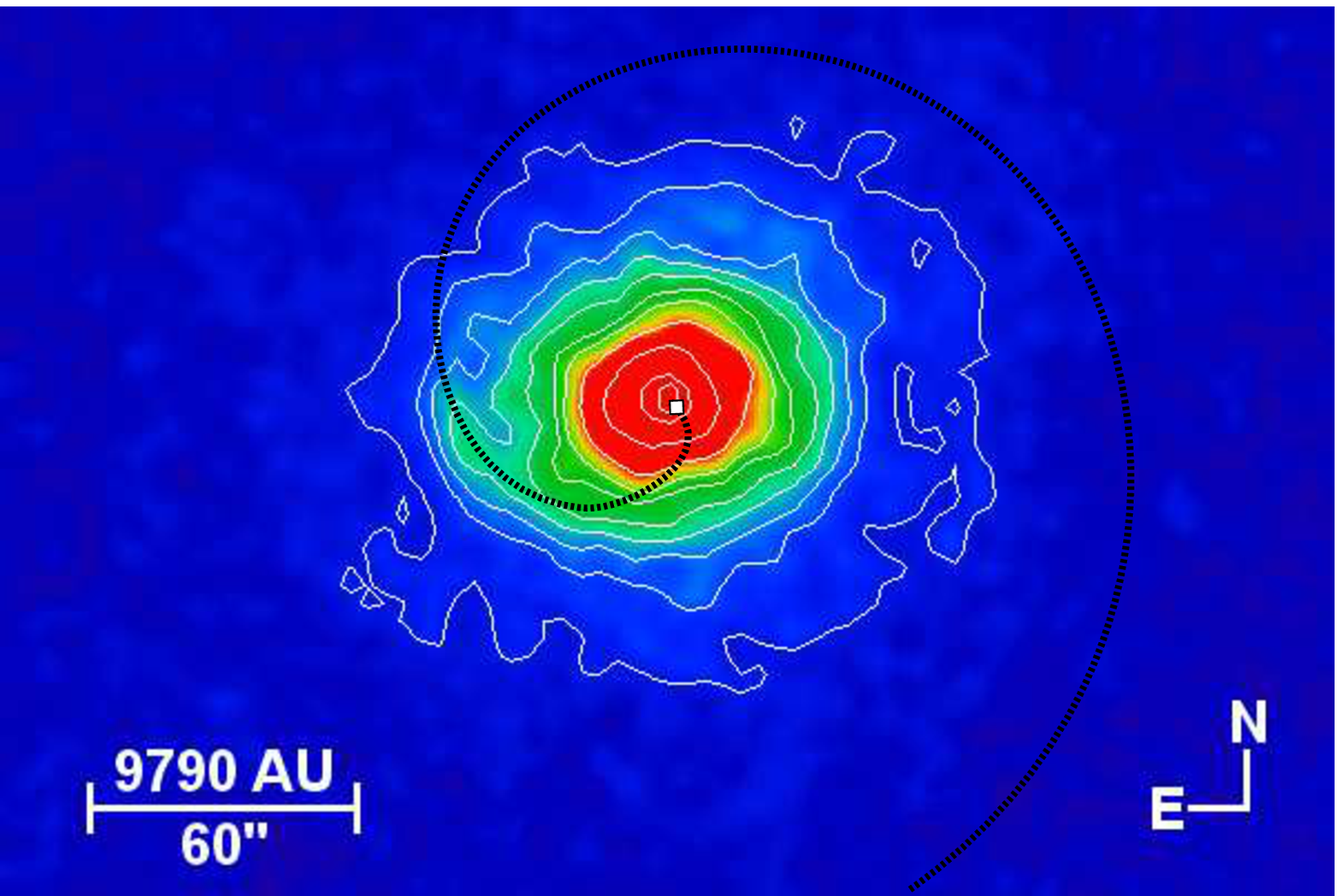} \\
   \vspace{0.5cm}
        \includegraphics[width=9cm]{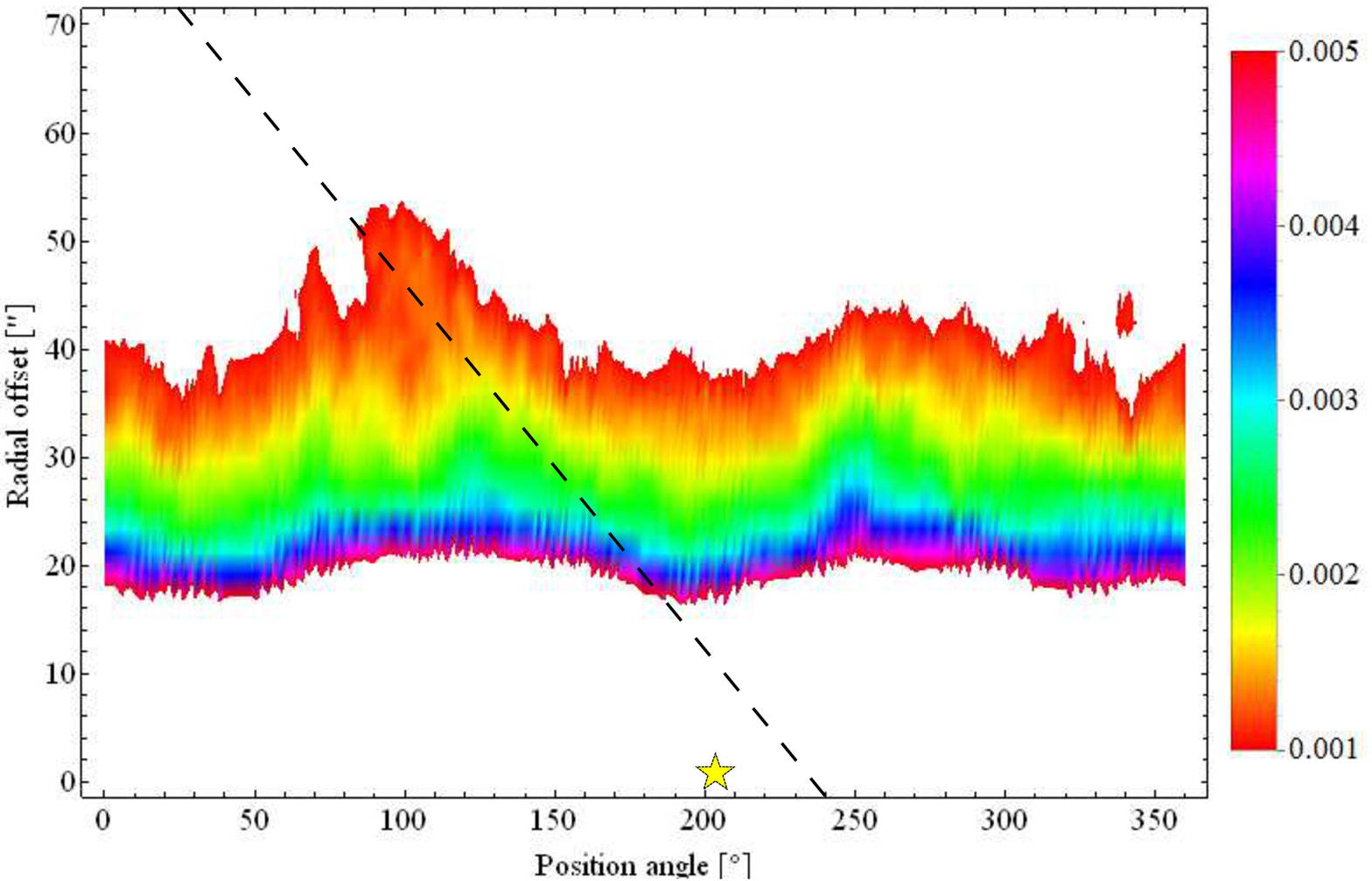} 
     \caption{\textit{Upper panel}: Contour plot of the \emph{Herschel}/PACS 70\,$\mu$m image of $\pi^1$~Gru over-plotted with an Archimedean spiral. The spiral spacing of $116\arcsec$ was derived from the wind velocity and the orbital period of the known G0V companion. \textit{Lower panel}: Polar-radial intensity profile of the same image. The position angle is measured north over east and the colour code is given in Jy\,arcsec$^{-2}$. The dashed line shows the same Archimedean spiral as in the upper panel. The (yellow) star illustrates the current position of the G0V companion.}   
\label{pi_gru_contours}
\end{figure}

In the upper panel of Fig~\ref{pi_gru_contours} we plot a spiral with $\rho = 116\arcsec$ over the PACS image that seems to match the observed arc well. However, this assumption has to be made
with care since the curvature of the arm allows many solutions. A unique solution is obtained when the start of the spiral coincides with the current position of the G0V star. The lower panel of Fig.~\ref{pi_gru_contours} displays the same scenario in a polar-radial diagram, which facilitates the illustration of the Archimedean spiral (dashed line). The current position of an anticlockwise orbiting companion is given by
\begin{equation}
\Phi_{\rm comp}=\Phi_{\rm arc}+\frac{2\pi R}{\rho}
,\end{equation}
where $R$ is the distance from a given part of the spiral to its origin, and $\Phi_{\rm arc}$ the PA of that given part\footnote{This is a simplification since the origin of the Archimedean spiral is the primary. For a binary system this wide, it is assumed,
however, that the accretion wake of the companion  causes the spiral, not the reflex motion of the primary. The spiral thus follows the involute of the circular orbit. Since the outcome is almost identical \citep{Kim2012b}, we keep the description of an Archimedean spiral.}. Assuming $R=52\arcsec-2\farcs8$, $\Phi_{\rm arc}=85\degr$, and $\rho=116\arcsec$ , the PA of the companion is $\approx240\degr$, which has to be compared with the observed PA $\Phi_{\rm comp}=203\degr$ (see Table~\ref{Tab:pi_gru_pos}). 

A spiral that matches both the slope of the arc and the position of the G0V star has an arm spacing $\rho=168\arcsec$. This value is about 45\% higher than that derived from the wind velocity of the primary, however, and the orbital period and this difference may originate from the uncertainties on these values.

\citet{Maercker2012} remarked that the arm separation of the spiral around R~Scl changed significantly during the past 1800\,yr. The outer (older) part of the spiral shows a larger separation than the inner (younger) part. For the authors, this indicates a modulation of the mass-loss rate by a factor of 30 caused by a thermal pulse. At the beginning of that phase, the wind velocity increased by 40\% and subsequently declined to the present-day value within 1200\,yr. This is measurable in the spiral-arm separation, which varied by this value. 

Such a thermal pulse presumably also occurred in $\pi^1$~Gru given its location at the tip of the AGB. Furthermore, \citet{Knapp1999} suggested that the mass-loss rate of $\pi^1$~Gru has increased in the past 1000 years to explain the presence of the CO disc. 

The second uncertainty on the arm separation $\rho$ stems from the assumption that the observed orbital separation of the $\pi^1$~Gru+G0V system ($\approx2\farcs8$) is de-projected, meaning that the binary orbit is seen face-on. For instance, to obtain an arm spacing of $168\arcsec$ with a constant wind velocity of 14.5\,km\,s$^{-1}$, an orbital period of 9000\,yr is required, which implies an inclination $i=46\degr$ of the orbit w.r.t. the plane of the sky. 

The appearance of inclined spirals was studied in the hydrodynamic simulations of \citet{Mastrodemos1999}, \citet{Mohamed2011}, and \citet{Kim2012b}. The authors found that the spiral shape is preserved up to an inclination angle of $\approx70\degr$ and then changes its appearance to broken concentric shells. Hence, a spiral pattern inclined by $46\degr$ would still be recognisable as such.

\subsection{Origin of the elliptical emission}
\label{Sec:elliptical}

A puzzling fact is the presence of the elliptical emission together with the spiral arc. While the arc can be explained by the G0V companion interacting with the primary's wind (see Sect.~\ref{Sec:arc}), this cannot be the case for the elliptical emission. 

An inclined disc structure around $\pi^1$~Gru was proposed by several CO and SiO line studies in the past 20 years \citep[e.g.][]{Sahai1992,Winters2003,Knapp1999,Chiu2006}. In all of the observations, the CO(2--1) profiles show an asymmetric, double-peaked structure and extended emission wings. On a corresponding map \citep{Chiu2006}, the envelope is elongated in the east-west direction with a size of $\approx40\arcsec$ [6530\,au] and a velocity gradient in the north-south direction.  

\citet{Knapp1999} interpreted these observations in terms of an expanding disc with a radius of 3340\,au [$20\arcsec$] inclined by $35\degr$ to the plane of the sky with the northern part of the disc tilted away from the observer. The same inclination was found from the axis ratio of the PACS far-IR emission along with the same orientation of the projected major axis, which assumes that both structures are identical, but visible on different scales. A similar conclusion was drawn by \citet{Ladjal2010b} from the APEX/LABoCa observations where the authors mention an inclination of the structure of about $70\degr$ (a probable misprint for $48\degr = \arccos[40/60]$).

According to \citet{Knapp1999}, the disc is produced by a constant mass-loss rate of $1.2\times10^{-6}$\,$M_{\sun}$\,yr$^{-1}$ expanding in the plane of the disc with a velocity of $13 \pm 2$\,km\,s$^{-1}$, in agreement with the velocity of the SiO(6--5) line. The expansion velocity derived from the CO lines increases steadily to 18\,km\,s$^{-1}$ towards the pole. In addition, a fast molecular wind (with velocities of at least 70\,km\,s$^{-1}$) is observed, which is most likely a continuation of the velocity increase towards the poles. This is in conflict with \citet{Sahai1992}, who interpreted the spatially-separated horn features in the CO lines as arising from a bipolar flow perpendicular to the disc. In the most recent study of the CO emission, \citet{Chiu2006} adhered to the \citet{Knapp1999} model and found, moreover, that the high-velocity outflow is oriented along PA=30$\degr$ -- 210$\degr$.

A natural explanation for the elliptical emission is that the wind of the AGB star does not propagate in a spherically symmetric fashion but is focused towards a plane. Three known mechanisms can account for this: (i) a fast differential internal rotation creates a gradient in the wind velocity between the equator of the star and its poles. However, only one AGB star is known to show a fast rotation \citep[V~Hya;][]{Barnbaum1995}, and generally slow rotation rates are found among white dwarf stars \citep{Kawaler1999}; (ii) a bipolar magnetic field causes the wind to become denser at the magnetic equator \citep{Matt2000}. Surface magnetic fields have been recently reported for some AGB stars \citep{Vlemmings2011,Ferreira2013,Lebre2014}, but there is no observational evidence for stellar magnetic fields to shape stellar AGB winds; (iii) the most common explanation is gravitational focusing of the AGB wind on the orbital plane of a companion. This scenario is supported by hydrodynamic simulations \citep{Mastrodemos1999,Kim2012c} and by observations \citep[e.g.][]{vanWinckel2009}.

If the elliptical emission is part of the primary's wind that is focused towards the orbital plane of the G0V companion, that star is orbiting {\it within} the disc. Since the high-velocity receding lobe is almost centred on the position of the secondary \citep[see Fig.~4 of][]{Chiu2006}, it is hard to imagine that the G0V companion does not produce any disturbance in the low-velocity CO disc. Therefore, the orientation of the CO disc might not be aligned with the orbital plane of the G0V companion.

The enormous separation of the system ($d>460$\,au) additionally makes it unlikely that the main-sequence companion is able to focus the AGB wind towards the orbital plane. \citet{Mastrodemos1999} used models with binary separations of 3.6\,au to 50.4\,au and $M_{\rm prim}/M_{\rm comp} = 0.75 - 6$ in their hydrodynamic simulations, but only the models up to 12\,au were able to form bipolar or elliptical circumstellar envelopes \citep[see Table~2 of][]{Mastrodemos1999}. The density contrast between the mass-accretion rate of the secondary and the mass-loss rate of the primary thereby defines the degree of focusing of the stellar wind and can be estimated after \citet{Morris1990} as
\begin{equation}
\alpha_{\rm foc} \equiv \frac{\dot{M}_{\rm acc}}{\dot{M}_{\rm prim}} = \left(\frac{G M_{\rm comp}}{d}\right)^2 \frac{1}{v_{\rm w}} \left(v_{\rm w}^2 + \frac{G(M_{\rm comp}+M_{\rm prim})}{d}\right)^{-3/2}.
\end{equation}
Assuming $M_{\rm prim} = 1.5$\,$M_\sun$, $M_{\rm comp} = 1.0$\,$M_\sun$, $v_{\rm wind} = 14.5$\,km\,s$^{-1}$, and $d=460$\,au, the focusing ratio is $\alpha_{\rm foc} =8 \times 10^{-5}$. According to \citet{Han1995}, strong and mild focusing is expressed by $\alpha_{\rm foc} > 0.1$, which is more than three orders of magnitude higher than the value found for the $\pi^1$~Gru+G0V system. The shaping agent of the elliptical CSE observed by CO and dust emission might therefore be another object that is located much closer to the AGB star than the G0V companion. This hypothesis was first expressed by \citet{Chiu2006} and is discussed here in Section~\ref{pi_gru_triple}.

If the elliptical emission is indeed an inclined disc that is not located in the orbital plane of the G0V companion, the question arises whether the mass accreted by the star is large enough to form the arc. Given the large system separation of more than 460\,au, the accretion rate on the companion is very low, even if the stellar AGB wind expands isotropically. The focusing ratio $\alpha_{\rm foc} =8 \times 10^{-5}$ is a factor of four lower than the lowest value simulated by \citet{Mastrodemos1999} in their model M9. But even at this low rate, a spiral pattern is forming. Nevertheless, the accretion rate of the G0V star is surely enhanced when the star moves through the disc, that is, in the region where the orbital plane and the disc intersect. Unfortunately, the orbital parameters are unknown because of the long period of the G0V companion. 

Another interpretation of the CSE ellipticity is that it represents a deformed asterosphere caused by the stellar wind interacting with the ISM. \citet{Ueta2006}, \citet{Jorissen2011}, and \citet{Cox2012} showed that fast-moving AGB stars can alter the wind bubble and produce a bow shock in the direction of the space motion at the interface of the wind and the ISM. For stars with a low space velocity, the CSE appears elliptical \citep{Weaver1977}. Adopting the long time-scale proper motion from the Tycho-2 catalogue (see Table~\ref{Tab:kinematic_pi_gru}), the direction of the space motion is at PA~$=103.1\degr\pm4.0\degr$, which means that
it is aligned with the major axis at PA~$\approx105\degr$. However, the velocity of the space motion is only $15.0\pm2.8$\,km\,s$^{-1}$, which is comparatively low to cause the elongation. \citet{Cox2012} nevertheless showed that even stars with a space velocity similar to $\pi^1$~Gru are able to form bow shocks (\object{R~Leo} and \object{UU~Aur}). The presence of the arc in the direction of the space motion, however, diminishes the possibility that the elliptical emission is shaped by the oncoming ISM.

\subsection{A hidden companion in the $\pi^1$~Gruis system?}
\label{pi_gru_triple}

As shown above, the elliptical emission around $\pi^1$~Gru cannot be shaped by the G0V companion given its distance and mass. Therefore, an object closer to the star might focus the primary wind on the orbital plane, making $\pi^1$~Gru a hierarchical triple system. Interestingly, \citet{Chiu2006} found a central cavity with a radius of 200\,au ($1\farcs2$) in their CO map. The authors note that it is large enough to host a putative close companion, but not the G0V companion orbiting the S star at an angular distance of $2\farcs8$. In the following sections, we discuss further indications for a second companion from astrometric and interferometric observations.

\subsubsection{$\Delta\mu$ behaviour and the Hipparcos Intermediate Astrometric Data}
\label{sect:hipparcos}
\begin{figure}[t!]
\centering
   \includegraphics[width=9cm]{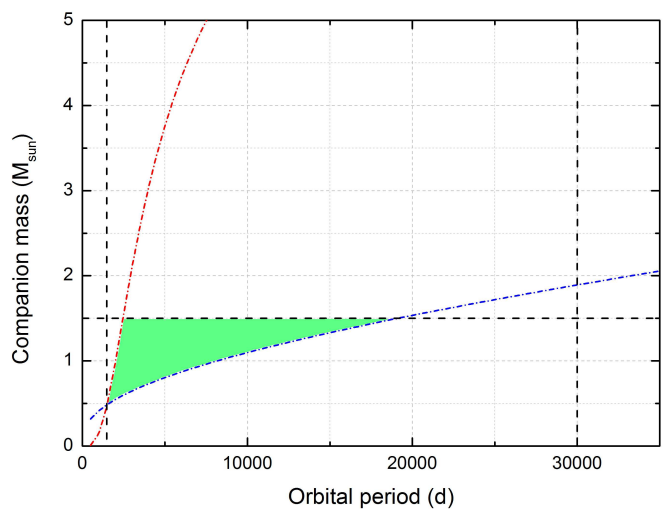} 
     \caption{Constraints on the orbital period and the companion mass $M_2$ set by the $\Delta\mu$ criterion (Eq.~\ref{Eq:constraint}; above the blue line) and by the Roche lobe criterion (Eq.~\ref{Eq:Roche}; below the red line) adopting a stellar radius of 420\,$R_{\sun}$ and a primary mass of 1.5\,$M_\sun$. The latter condition imposes the companion to be less massive than 1.5\,$M_\sun$ (below the dashed horizontal line). The ability to detect a $\Delta\mu$ binary moreover imposes $1500 < P {\rm (days)} < 30\,000$ (between both vertical dashed lines). The admissible region is enclosed within these boundaries (green area).}
     \label{Fig:M2_P_constraints}
\end{figure}
$\pi^1$~Gru is found to be a $\Delta\mu$ binary \citep{Makarov2005,Frankowski2007}, meaning that its long-term proper motion \citep[Tycho-2;][]{Hog2000} is different from its short-term proper motion \citep[Hipparcos;][]{vanLeeuwen2007}, because the latter is altered by the orbital motion while the orbital motion averages out on the long-term proper motion (see Table~\ref{Tab:kinematic_pi_gru} and also Appendix~\ref{Sect:space_motion}). 

In the following we investigate whether the $\Delta\mu$ binary nature of $\pi^1$~Gru can be caused by the G0V companion separated by more than 460\,au. The observed proper motion discrepancy is derived as 
\begin{equation}
\begin{split}
\Delta\mu& = \left[\left(\mu_{\alpha*}(\mathrm{HIP})-\mu_{\alpha*}(\mathrm{TYC2})\right)^2 + \left(\mu_{\delta}(\mathrm{HIP})-\mu_{\delta}(\mathrm{TYC2})\right)^2\right]^{1/2} \\
 & =7.4\pm1.4{\rm \,mas\,yr^{-1}}.
\end{split}
\end{equation}
\citet{Frankowski2007} have shown, thanks to a comparison with known
spectroscopic binaries from the {\it Ninth Catalogue of Spectroscopic
Binary Orbits} \citep[S$_{\rm B^9}$;][]{Pourbaix2004}, that the binaries
detectable by the $\Delta\mu$ approach must have orbital periods in the range of 1500 to 30\,000 days. This already implies that the G0V companion with an orbital period of $\approx2.2\times10^6$ days can hardly account for the proper motion variation. For a more detailed analysis, \citet{Makarov2005} showed that $\Delta\mu$ is related to the orbital parameters in the following way:
\begin{equation}
\label{Eq:Deltamu}
\Delta\mu \le \frac{2\pi\varpi R_0 M_2}{(M_1+M_2)^{2/3}\;P^{1/3}}, 
\end{equation}
where $M_1$ and $M_2$ are the primary and secondary masses, $\varpi$ is the parallax, $P$ is the orbital period, and
$R_0$ is a time-dependent orbital phase term,
\begin{equation}
R_0 = \left(\frac{1+e \cos E}{1 - e \cos E}\right)^{1/2},
\end{equation}
where $e$ is the orbital eccentricity and $E$ the eccentric anomaly. In
the following, $R_0 = 1$ is assumed, equivalent to a circular orbit. 
\begin{figure}[t!]
\centering
   \includegraphics[width=9cm]{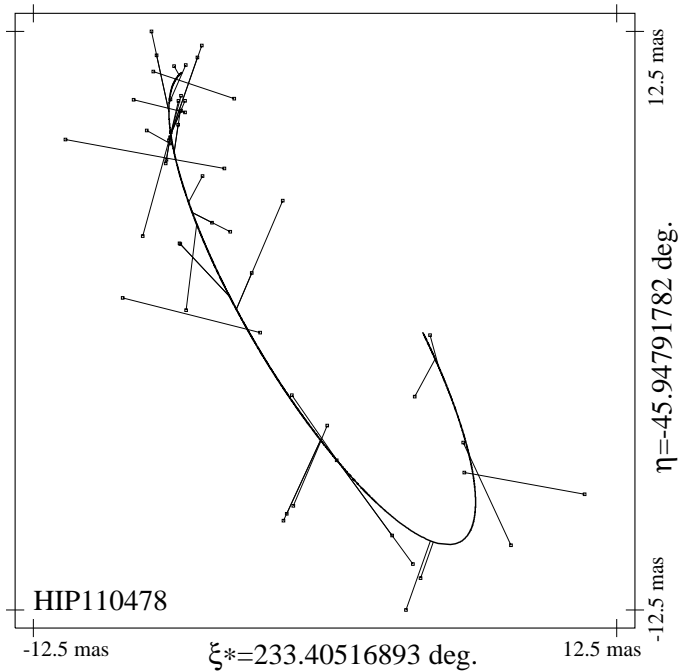} 
     \caption{Orbital arc derived from the analysis of the Hipparcos Intermediate Astrometric Data of $\pi^1$~Gru (see last entry in Table~\protect\ref{Tab:orbit}). The Hipparcos observations are 1D data, the bars therefore indicate the possible position of the photocentre perpendicular to the orbital segment.}
     \label{pi_gru_HIP}
\end{figure}

\begin{table}
\caption{Kinematic data of $\pi^1$~Gru from the Hipparcos \citep{vanLeeuwen2007} and Tycho-2 \citep{Hog2000} catalogues. $v_{\rm LSR}$ is the velocity of the star corrected for the solar motion, $i$ and PA the inclination to the sky plane and position angle of the space motion.}
\label{Tab:kinematic_pi_gru}
\begin{center}
\begin{tabular}{lrr}
\hline \hline
 & Hipparcos & Tycho-2 \\
\hline
$\mu_{\alpha*}$ (mas/yr) & $28.48\pm0.94$ & $33.4 \pm1.1$  \\
$\mu_\delta$ (mas/yr) & $-12.14\pm0.60$ & $-17.7\pm1.2$  \\
$v_{\rm LSR}$ (km/s) & $11.7\pm2.5$ & $15.0\pm2.8$  \\
$i$ ($\degr$) & $-40.4\pm15.0$ & $-30.4\pm13.2$  \\
PA ($\degr$) & $81.3\pm2.9$ & $103.1\pm4.0$  \\
\hline
\end{tabular}
\tablefoot{The parallax $\varpi = 6.13\pm0.76$\,mas from \citet{vanLeeuwen2007}, the radial velocity $V_{\rm r} = -5.7\pm1.5$\,km\,s$^{-1}$ from \citet{2000A&AS..145...51V}, and the solar motion vector $(U,V,W)_{\sun}=(8.50\pm0.29,13.38\pm0.43,6.49\pm0.26)$ from \citet{Coskunoglu2011} were used to derive $v_{\rm LSR}$, $i$, and PA.}
\end{center}
\end{table}

The equality in Eq.~\ref{Eq:Deltamu} corresponds to an orbit seen
face-on, thus all our estimations in the remainder of this section correspond to lower bounds. With $\varpi=6.13$\,mas \citep{vanLeeuwen2007} and $\Delta\mu=7.4$\,mas\,yr$^{-1}$,  Eq.~\ref{Eq:Deltamu} yields
\begin{equation}
\label{Eq:constraint}
0.192 P^{1/3} \le \frac{M_2}{(M_1+M_2)^{2/3}}.
\end{equation}
Another constraint on the orbital period comes from the Roche lobe radius $R_1$ \citep{Paczynski1971}. Since
$\pi^1$~Gru appears to be close to the tip of the AGB \citep[][and Sect.~\ref{Sect:properties}]{vanEck1998}, 
its large radius limits the admissible orbital separation and period
\begin{equation}
R_1 \le a \left(0.38 + 0.2\log\frac{M_1}{M_2} \right).
\end{equation}
By substituting $a$ from the third Kepler law\begin{equation}
P^2=\frac{4\pi^2}{G(M_1+M_2)}a^3
,\end{equation}
one obtains
\begin{equation}
R_1 \le P^{2/3} \left(\frac{G(M_1 + M_2)}{4\pi^2}\right)^{1/3} \left(0.38 + 0.2\log\frac{M_1}{M_2}\right),
\label{Eq:Roche}
\end{equation}
where $G$ is the gravitational constant. Eqs.~\ref{Eq:constraint} and \ref{Eq:Roche} allow us to restrict the range of possible values for $P$ and $M_2$ (Fig.~\ref{Fig:M2_P_constraints}) if we adopt 420\,$R_{\sun}$ for the radius of $\pi^1$~Gru (12\,mas at 163\,pc) as derived by \citet{Cruzalebes2006} from VLTI/MIDI observations, and $M_1=1.5$\,$M_\sun$ for the primary mass from its location in the Hertzsprung-Russell diagram \citep{vanEck1998}. Fig.~\ref{Fig:M2_P_constraints} presents these constraints in graphical form.

\begin{table}
\caption[]{\label{Tab:orbit}
Possible orbital solutions obtained from the analysis of the 55 Hipparcos Intermediate Astrometric Data of $\pi^1$ Gru (HIP~110478), by scanning a $P - e$ grid.  $F2$ is the goodness-of-fit, as defined by Eq.~\ref{Eq:GoF}.
}
\begin{tabular}{lllllll}
\hline\hline
$e$ & $P$  & $\chi^2$ & $F2$ & $\varpi$ & $\mu_{\alpha*}$ & $\mu_{\delta}$ \\
    & (yr) &           &     &  (mas)   & (mas yr$^{-1}$)  & (mas yr$^{-1}$) \\
\hline
0.5 & 8.3 & 60.45 & 1.74 & 6.55 & 26.1 & -19.1 \\
0.5 & 4.6 & 59.28 & 1.64 & 6.53 & 26.4 & -16.4 \\
0.7 & 4.6 & 56.74 & 1.42 & 6.55 & 27.3 & -17.8 \\
0.9 & 9.7 & 55.66 & 1.32 & 6.55 & 34.3 & -21.9 \\
0.9 & 4.6 & 54.01 & 1.17 & 6.66 & 28.9 & -17.9 \\
0.9 & 6.3 & 53.18 & 1.09 & 6.68 & 31.2 & -18.8 \\
\hline\\
\end{tabular}
\end{table}

We thus conclude from this simple analysis that the G0V companion separated by at least 460\,au with an orbital period of several thousand years cannot be the cause of the $\Delta\mu$ binary. A close second companion is required instead. An analysis of the Hipparcos IAD was performed along the method outlined by \citet{Pourbaix2000}\footnote{except for the fact that the condition imposing a positive parallax has since been found to be inappropriate and has been lifted in recent applications of the \citet{Pourbaix2000} method.} and especially by \citet{Jorissen2004}, to search for the possible presence of a close binary companion in the Hipparcos IAD {\it without any a priori knowledge of the orbital elements.} Basically, the routine scans a grid in eccentricity -- period and searches for the best possible solution (in terms of $\chi^2$ value) by including orbital motion at each grid point.  Satisfactory solutions (i.e., with $\chi^2$ in the range of 53 to 60, because 55 data points are available, or goodness-of-fit\footnote{If $\chi^2$ does follow a chi-square distribution with $\nu$ degrees of freedom, the goodness-of-fit  follows a $N(0, 1)$-distribution irrespective of $\nu$ \protect\citep[see e.g.,][]{Pasquato2011}:
\begin{equation}
F2=\sqrt{\frac{9\nu}{2}}\left[\sqrt[3]{\frac{\chi^2}{\nu}}+\frac{2}{9\nu}-1\right].
\label{Eq:GoF}
\end{equation}
}
values between 1.0 and 1.8) are obtained for eccentricities higher than 0.5 and orbital periods in the range of 5 to 11\,yr. Although the available data do not allow to fully constrain the orbit, confidence in the orbital solutions obtained from the Hipparcos IAD is bolstered because the proper motion derived from the analysis of the Hipparcos data now agrees with the long-term Tycho-2 proper motion. Possible solutions are listed in Table~\ref{Tab:orbit}. The orbital arc corresponding to the best fitting among these possible solutions (last entry in Table~\ref{Tab:orbit}) is presented in Fig.~\ref{pi_gru_HIP}. Moreover, the favoured orbital periods (4.6--9.7\,yr, or 1680--3540\,d) are in the range considered to be likely from the analysis of Eqs.~\ref{Eq:constraint} and \ref{Eq:Roche} (4.1--52.3\,yr). According to Fig.~\ref{Fig:M2_P_constraints}, periods as short as 4.6\,yr are only marginally possible, but solutions with orbital periods around 6\,yr or longer are perfectly admissible and imply masses for the companion in the range of 0.5 to 1.5\,$M_\sun$ that would correspond to spectral types K9 to F3 on the main sequence. 

A system similar to $\pi^1$~Gru + close companion was modelled by \citet{Mastrodemos1999} in their models M10 (1\,$M_\sun$ companion) and M17 (0.5\,$M_\sun$ companion). The radius of the primary ($R_{\rm p}=452.4$\,$R_\sun$) and the separation of the system ($a=6.3$\,au) are the same for both models and resemble the results obtained from interferometric observations \citep{Sacuto2008} and the $\Delta\mu$ estimate above. 

In model M10, the wind morphology is indeed collimated and oblate due to the gravitational force of the companion on the spherical wind of the AGB star. It is thus conceivable that a close 1\,$M_\sun$ companion is the cause for the disc observed in the CO and dust emission. A lower-mass companion, as in M17, however, prevents this behaviour and a well-defined spiral pattern occurs instead. The size of the spiral pattern can be evaluated with Eq.~\ref{arm_separation}, and results in $\rho\approx0\farcs2$ ($\approx 33$\,au). This is much smaller than the $1\arcsec$ pixel size of the $70\,\mu$m PACS image, which does not allow us to favour either of the models.

\subsubsection{Interferometric observations}
\label{Sect:AMBER_discussion}
The angular resolution of VLTI/AMBER is perfect to investigate the deformation of the envelope induced by a close secondary companion.
\begin{figure}[t!]
\centering
   \includegraphics[width=0.5\textwidth, angle=90,bb = 27 303 562 806]{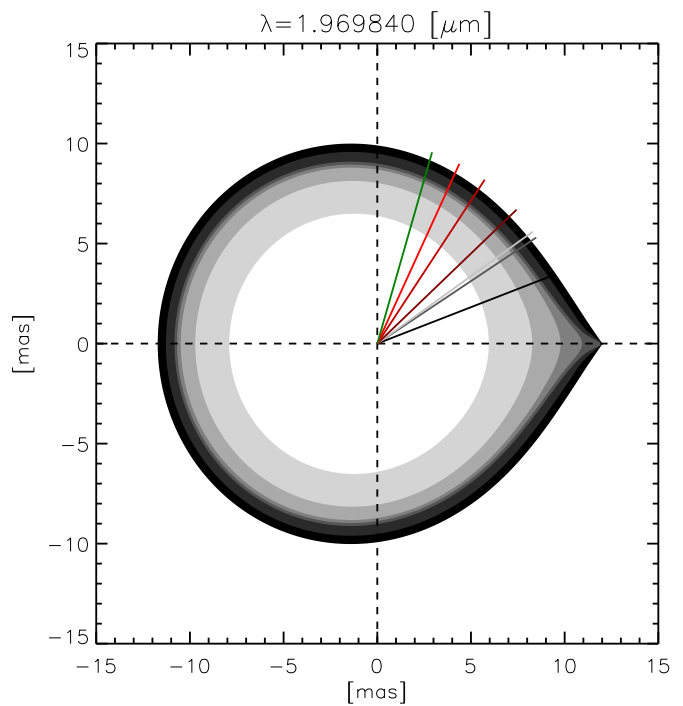} 
     \caption{Intensity map of the Roche lobe model for one of the AMBER low-resolution spectral channels. The lines represent the direction of the projected baselines used to simulate the interferometric data. Since the location of the close companion and the orientation of the orbital plane are unknown, the depicted Roche lobe orientation illustrates only one possible solution.}
     \label{Fig:roche-map}
\end{figure}

The usual first step for interpreting interferometric observations with limited $uv$-coverage, like the AMBER data presented here, is the comparison with geometric models.
For this purpose we made use of the software GEM-FIND described by \cite{klotz2012}. It is sufficient here to report that none of the geometric toy models could fit the data in a satisfactory way. Given the presence of closure-phase signatures (i.e. asymmetric structures), 1D model atmospheres cannot reproduce the observations either. Therefore, we decided to switch to a more realistic physical model. The presence of a binary companion very close to the primary would trigger a tidal deformation of the primary star. If the primary is close to filling its Roche lobe (and this possibility is not excluded by the results presented in Sect.~\ref{sect:hipparcos} since Roche lobe fitting giants are located along the left curve of Fig.~\ref{Fig:M2_P_constraints}), the shape will resemble that of a pear, as shown in Fig.~\ref{Fig:roche-map}. This kind of geometry will produce a signature in the closure phase.

\cite{siopis2012} developed the software package Gaia 
Eclipsing System Simulator and Solver (GESSS) with the primary aim of modelling the light curve of eclipsing binaries for the Gaia survey. This tool is very flexible and can also be used to model other binary configurations. The code was recently adapted to extract interferometric observables \citep{Paladini2014}.
\begin{figure*}[t!]
\centering
\includegraphics[width=14cm, bb = 23 20 559 804]{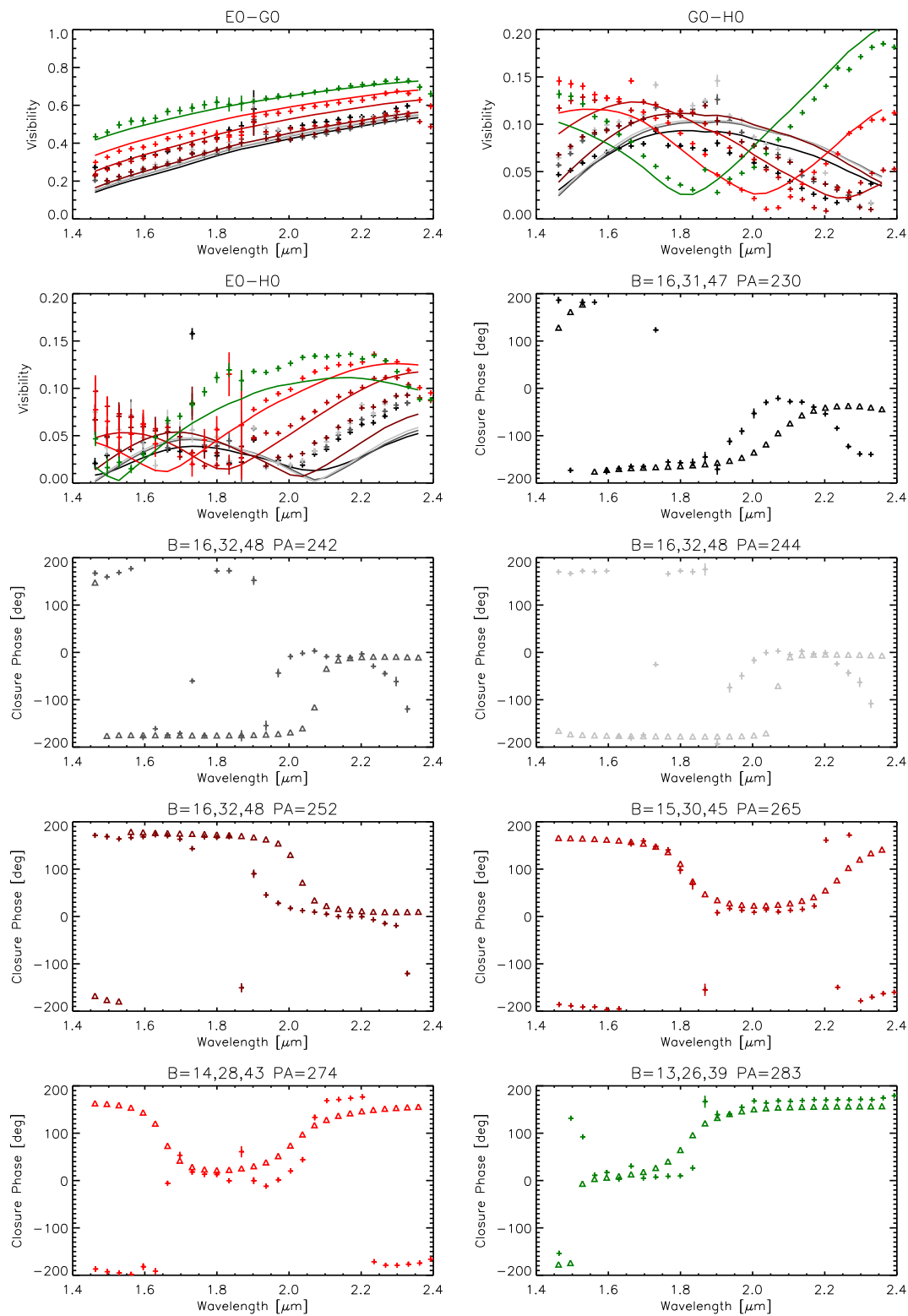} 
     \caption{Comparison between AMBER data (crosses) and the synthetic Roche lobe observations (full line for the upper three panels, triangles for the others; see Fig.~\ref{Fig:roche-map}). The first three panels depict the visibilities of the E0-G0-H0 configuration (see Table~\ref{tab:amber_obs}). The other panels show the closure phases produced by the triplet of baselines B at position angle PA. }
     \label{Fig:binarymodel-interf}
\end{figure*}

For our experiment, we used as starting point a MARCS model atmosphere \citep{gustafsson2008} with stellar parameters  $T_{\rm eff} = 3200$~K, $\log g = 0.3$, and solar metallicity. The Gaia eclipsing binary software identifies the stellar surface with Roche equipotentials, which are numerically computed for each component of the binary system using a dense mesh of points. This mesh defines a scalar field of intensities (calculated from the MARCS model that incorporates the limb darkening), which is then linearly interpolated to produce a synthetic 1020-by-1020-pixel image of the system. We produced a set of 27 intensity maps spread across the $H$ and $K$ bands, with the spectral resolution of AMBER ($R = 30$). This set of images was produced for two Roche lobe models, one with mass ratio 1/3, and one with mass ratio 1. As we do not know the orientation of the orbital plane, we assumed for simplification that for both models the companion is currently located at one of the orbital nodes, that is, at the intersection of the orbital plane with the plane of the sky. Thus, the Roche lobe is seen face-on. An example of the intensity maps in one of the AMBER low-resolution spectral channels is shown in Fig.~\ref{Fig:roche-map}. 
\begin{figure*}[t!]
\centering
\includegraphics[width=18cm]{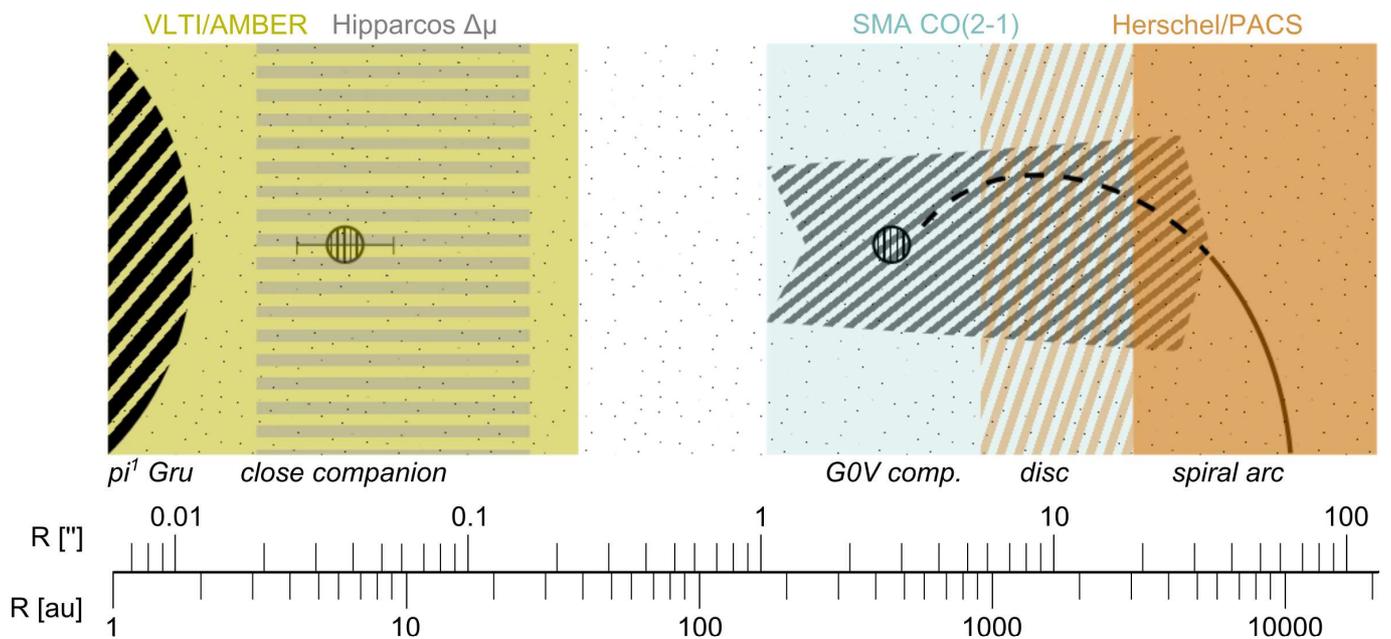} 
\caption{Illustration of the surroundings of $\pi^1$~Gru including the close companion, the disc, the G0V companion, and the spiral arc. The coloured areas represent the field of view or detection range of the used instrument: olive represents the VLTI/AMBER range, horizontal grey lines the Hipparcos $\Delta\mu$ constraint, light blue the SMA CO(2-1) observations, and dark orange the partly overlapping \emph{Herschel}/PACS field of view.}
\label{Fig:pi_gru_illu}     
\end{figure*}

The orientation of the $\pi^1$~Gru system on the sky with respect to the projected baselines is unknown. In principle, the full parameter space of azimuthal {\it and} polar angles describing the orientation of the system on the plane of the sky has to be explored. Together with the position angle of the baselines, similar images to that of Fig.~\ref{Fig:roche-map} have to be generated for all these possible orientations to compute the corresponding visibilities and closure phases. This effort is beyond the scope of this paper, and we here restrict ourselves to showing that the Roche lobe model may yield closure-phase variations qualitatively similar to those observed. We chose the system orientation that yielded the highest asymmetry (shown in Fig.~\ref{Fig:roche-map}), and searched for the baseline position angle that yielded closure phases that varied in the same way
as the observations. Starting from a first baseline with a position angle of either -60, -40, 0, 10, 20, or 30 degrees, we extracted visibilities and closure phases for the same baseline pattern as displayed in Fig.~\ref{fig:uv}.
 
In Fig.~\ref{Fig:roche-map}, we show the orientation of the baseline pattern of Fig.~\ref{fig:uv} needed to obtain the model curves displayed in Fig.~\ref{Fig:binarymodel-interf}. These model curves should by no means be considered as best fits, but they illustrate the good prospects offered by the Roche model. Nevertheless, it has to be kept in mind that not only the Roche lobe geometry can cause non-zero signatures in the closure phase, but so can stellar spots or flares \citep[e.g.;][]{Chiavassa2010,Wittkowski2011}.

The Roche lobe model suggests that a solution with the smallest orbital separation (red line in Fig.~\ref{Fig:M2_P_constraints}, corresponding to Roche lobe filling) is consistent with both the AMBER and Hipparcos data. The AMBER data do not allow us to unambiguously select one among the possible Hipparcos solutions (Table~\ref{Tab:orbit}), but they offer confidence for the hypothesis of a close companion in the $\pi^1$~Gru system.

As a summary, the whole scenario including the close companion, the disc, the G0V companion, and the spiral arc are illustrated in Fig.~\ref{Fig:pi_gru_illu}. The coloured areas represent the field of view or detection range of the respective observing facility. The horizontal grey lines around the close companion indicate the range of acceptable orbital separations using the Hipparcos $\Delta\mu$ constraint. This is refined to the region indicated by the ``error bar'' which is based on the results from the Hipparcos IAD fitting (see Table~\ref{Tab:orbit}).

\section{Conclusions and summary}
\label{conclusions}

We have analysed the CSE of the highly evolved AGB star $\pi^1$~Gru based on \emph{Herschel} observations at 70\,$\mu$m and 160\,$\mu$m as part of the MESS sample. The images show an asymmetric stellar wind morphology with two main features, namely an elliptical CSE and an arc east of the star. The arc emerges $38\arcsec$ away from the star along the major axis of the ellipse and is curved towards the north-east before it becomes too faint. The arc is most likely a small part of an Archimedean spiral caused by the interaction of the stellar AGB wind with a companion. $\pi^1$~Gru has a physically related G0V companion that has been known for over a century and is separated by 460\,au ($2\farcs8$) from the primary. We were able to fit the observations with a spiral given the properties of the G0V star. For a perfect match, however, the wind velocity has to be adjusted to higher values, as was suggested by \citet{Knapp1999} to explain their CO observations. 

The second far-IR feature, the elliptical CSE, stretches over $72\arcsec \times 60\arcsec$ [$11750\times9790$\,au] and represents the dusty counterpart of CO emission with the same axis ratio, but it is slightly smaller \citep{Knapp1999,Chiu2006,Sahai1992}. All of the authors interpreted their findings as a disc structure inclined by 35$\degr$ to the plane of the sky, which is also
supported by the axis ratio of the far-IR emission. In the CO map by \citet{Chiu2006} the disc has an inner radius of $1\farcs2$ and an outer radius of $\approx20\arcsec$. Based on the focusing ratio, it can be ruled out that the known G0V companion focuses the AGB wind towards the orbital plane given the enormous separation of the system. Furthermore, the G0V companion would be orbiting within the disc without causing any observable disturbance. Because
of this, we assumed that the disc is not located in the orbital plane of the G0V companion and followed the hypothesis of \citet{Chiu2006} that $\pi^1$~Gru may have a close second companion. 

We found support for this assumption in several observations. $\pi^1$~Gru is a known $\Delta\mu$ binary, meaning that its long-term and short-term proper motions are significantly different \citep{Makarov2005,Frankowski2007}. An analysis of the Hipparcos \citep{vanLeeuwen2007} and  Tycho-2 \citep{Hog2000} data eliminates the G0V companion as the source of disturbance and suggests a 0.5--1.5\,$M_\sun$ companion with an orbital period in the range of 4--50\,yr. This result is strengthened by the Hipparcos IAD, which reveal an orbital motion of the photocentre of $\pi^1$~Gru. This motion is best fitted by a highly eccentric orbit with a period of 4.6--9.7\,yr. \citet{Mastrodemos1999} used this configuration in their hydrodynamic simulations where the companion was indeed able to focus the primary's wind towards the orbital plane. 

To obtain direct indications for the close companion we used archival data from VLTI/AMBER that show closure-phase signatures. Although the interferomeric observations can be qualitatively reproduced by a Roche lobe model, we cannot exclude that a more complex model including the presence of spots or flares will also be able to reproduce these observations. The main restriction comes form the $uv$-plane coverage. A VLTI/PIONIER imaging programme would probably help to break the degeneracy in the currently available data. Based on the current observational status, $\pi^1$~Gru is most likely a hierarchical triple system in which the close companion shapes the disc observed in CO and dust emission, while the (previously known) G0V companion is located farther outside and causes the spiral arc visible in the \emph{Herschel}/PACS images.

\begin{acknowledgements}
We thank B.~Mason (USNO) for providing the individual position measurements of the companion of $\pi^1$~Gru from the WDS catalogue and the variable star observations from the AAVSO International Database contributed by observers worldwide and used in this research. We also thank I.~Platais for advice with the proper motion evaluation. This work was supported in part by the Belgian Federal Science Policy Office via the PRODEX Programme of ESA. AM and FK acknowledge funding by the Austrian Research Promotion Agency FFG under project number FA 538019, RO under project number I163-N16. Furthermore, AM acknowledge funding by the Abschlussstipendium 2014 of the University of Vienna. This work used data from the ESO archive (Programmes 076.D-0624, 077.D-0620, 078.D-0122, 079.D-0138, 080.D-0076, 187.D-0924) and from the HST archive (Programs ID 3603 and 10185). This research has made use of the \textit{AMBER data reduction package} of the
Jean-Marie Mariotti Center\footnote{Available at \url{http://www.jmmc.fr/amberdrs}}.
\end{acknowledgements}

\bibliographystyle{aa}
\bibliography{biblio}

\begin{appendix}

\section{Positional data}

\subsection{$\pi^1$~Gru}
\label{Sect:space_motion}

\begin{figure}[t!]
\centering
\includegraphics[width=8cm]{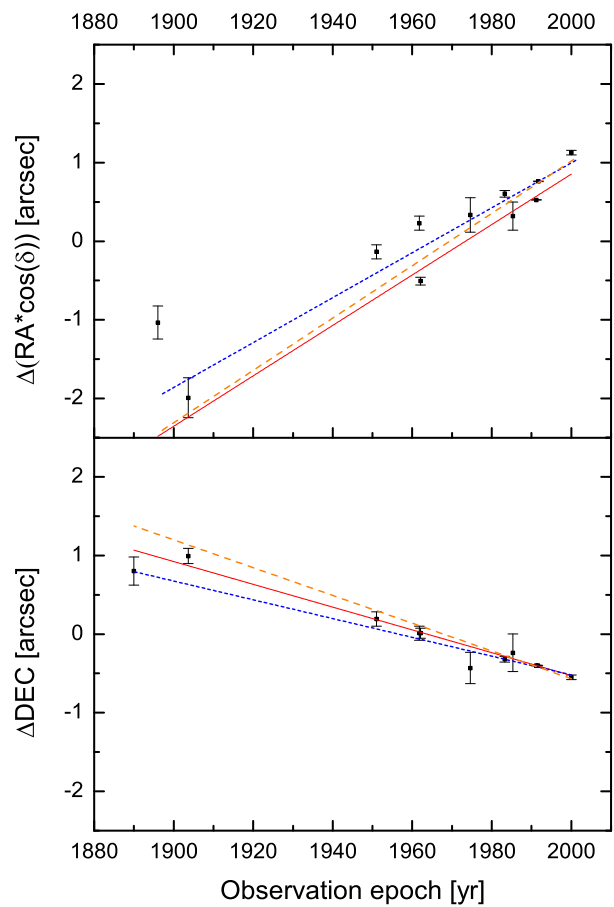} 
\caption{Positions of $\pi^1$~Gru from catalogues spanning more than 100 years. 
The solid red line corresponds to a linear fit
through the positions, weighted by the inverse square of the position uncertainties. 
The dashed blue line shows the positions extrapolated 
according to the Hipparcos proper motion, while
the dashed orange line represents the same for the Tycho-2 proper motion.}
\label{Fig:pi_gru_positions}     
\end{figure}

The difference between the long- and short-term proper motions is illustrated in Fig.~\ref{Fig:pi_gru_positions}, which shows the positions of $\pi^1$~Gru in the last 100 years from various catalogues (see Table~\ref{Tab:pi_gru_cata} for a list of these catalogue positions). The position at the observation epoch of a given catalogue, if not directly given by the {\it Vizier} database at the {\it Centre de Donn\'ees Stellaires} (Strasbourg), has been derived by applying the catalogue proper motion to the listed J2000 epoch position. Care has been exercised to ensure consistency between the equinox of the proper motion and the position. Our own estimate for the proper motion is obtained from a linear fit on all these positions, weighted by the inverse square of the uncertainty on the position. 

The resulting space velocity $v_{\rm LSR}$ \citep[corrected for the solar motion vector $(U,V,W)_{\sun}=(8.50\pm0.29,13.38\pm0.43,6.49\pm0.26)$;][]{Coskunoglu2011} based on these proper motions and the radial velocity derived by \citet{2000A&AS..145...51V} is listed in Table~\ref{Tab:kinematic_pi_gru}. Since the long-term Tycho-2 proper motion appears slightly more precise than the one we derived from the positions displayed in Fig.~\ref{Fig:pi_gru_positions}, the Tycho-2 proper motion is adopted in the remainder of this paper.

\begin{table*}[t!]
\caption{Positional data of $\pi^1$~Gru from various catalogues over the past century. All positions are given in the J2000.0 system.}
\label{Tab:pi_gru_cata}
\begin{tabular}{l|rrrrrrrrrrrrr}
\hline \hline
 &  SAO$^{(1)}$ & AC 2000.2$^{(2)}$ & PPM$^{(3)}$ & YZC$^{(4)}$ & CPC-2$^{(5)}$ & SHCB$^{(6)}$  \\
\hline 
Mean epoch of RA & 1896 & 1903.7 & 1950.9 & 1961.8 & 1962.1 & 1974.6  \\
Mean epoch of DEC & 1890 & 1903.7 & 1951.0 & 1961.8 & 1962.1 & 1974.6 \\
$\mu_{\alpha*}$ (mas\,yr$^{-1}$) & $9\pm11$ & - & $38.2\pm4$ & 39 & $38.6 \pm 3$ &  - \\
$\mu_{\delta}$ (mas\,yr$^{-1}$) & $-7\pm8$ & - & $-17\pm4$ & -22 & $-13 \pm 3$ & -  \\
RA at obs. epoch [$\degr$] & 335.682791 & 335.682854 & 335.683458 & 335.683616 & 335.683791 & 335.684319 \\
Error in RA [$10^{-5}$ $\degr$] & 5.833 & 7.056 & 2.5 & - & 1.361 & 6.111 \\
DEC at obs. epoch [$\degr$] & -45.947583 & -45.947530 & -45.947752 & -45.947802 & -45.947802 & -45.947926 \\
Error in DEC [$10^{-5}$ $\degr$] & 5 & 2.694 & 2.5 & - & 1.778 & 5.556 \\
\hline \hline
 & SPM4$^{(7)}$ & FOCAT-S$^{(8)}$ & Tycho-2$^{(9)}$ & PPMXL$^{(10)}$ & USNO-B$^{(11)}$  \\
\hline
Mean epoch of RA [yr] & 1983.3 & 1985.3 & 1991.4 & 1991.2 & 2000.0 \\
Mean epoch of DEC [yr] & 1983.3 & 1985.3 & 1991.4 & 1991.2 & 2000.0 \\
$\mu_{\alpha*}$ [mas\,yr$^{-1}$] & $26.9\pm4.8$ & $32 \pm 1$ & $33.4 \pm 1.1$ & $35.3 \pm 2$ &  32 \\
$\mu_{\delta}$ [mas\,yr$^{-1}$] & $-20.2\pm4.3$ & $-12 \pm 4$ & $-17.7 \pm 1.2$ & $-16.3\pm2$ & $-18$ \\
RA at obs. epoch [$\degr$] & 335.684025 & 335.684000 & 335.684094 & 335.684177 & 335.684203 \\
Error in RA [$10^{-5}$ $\degr$] & 1.183 & 5 & 0.111 & 0.056 & - \\ 
DEC at obs. epoch [$\degr$] & -45.947895 & -45.947872 & -45.947916 & -45.947915 & -45.947959 \\
Error in DEC [$10^{-5}$ $\degr$] & 1.025 & 6.667 & 0.111 & 0.056 & - \\
\hline
\end{tabular}
\tablebib{ (1): Smithsonian Astrophysical Observatory Star Catalog \citep{SAO1966}, (2): The Astrographic Catalogue on the Hipparcos System \citep{Urban2001}, (3): Positions and Proper Motions - South \citep{Bastian1993}, (4): Yale Zone Catalogues Integrated \citep{Fallon1983}, (5): Cape Photographic Catalogue 2 \citep{Nicholson1984}, (6): Southern Hemisphere Catalogue of Bordeaux \citep{Rousseau1996}, (7): Yale/San Juan Southern Proper Motion Catalog 4 \citep{Girard2011}, (8): Pulkovo photographic Catalogue of Southern Hemisphere \citep{Bystrov1994}, (9): The Tycho-2 Catalogue of the 2.5 Million Brightest Stars \citep{Hog2000}, (10): The PPMXL catalog of positions and proper motions on the ICRS \citep{Roeser2010}, (11): The USNO-B Catalog \citep{Monet2003}} 
\end{table*}

\subsection{G0V companion}
\label{App:Pos.G0V}

Table~\ref{Tab:pi_gru_pos} shows the separations and position angles of the G0V companion. The observations do not show significant changes of the companion's position over a timespan of more than 100 years. However, in 1989.86, the system was observed by \citet{Sahai1992}, who reported the position of the companion with a projected separation of $2\farcs45$ at PA=200.4$\degr$, clearly different from previous observations. The author remarks that such a fast motion for an object located at least $460$\,au away from the primary would imply unrealistically large stellar masses. Moreover, given the large uncertainties induced by the $0\farcs9$ seeing that prevailed during the observations, no reliable conclusions regarding the orbital period could be drawn by \citet{Sahai1992}. 

\begin{table}
\caption{Separations and position angles of the G0V companion of $\pi^1$~Gru, from the 
Washington Double Star Observations
catalogue (courtesy of B.~Mason).}
\label{Tab:pi_gru_pos}
\begin{tabular}{lllllll}
\hline \hline
Obs. Epoch &  PA  & d & Reference \\
(yr) & ($\degr$) & ($\arcsec$) & \\
\hline
1896.8 & 190 & 2.5  & \citet{Innes1897} \\
1900.75 & 201.6 & 3.04 & \citet{Innes1905} \\
1900.76 & 201.4 & 2.70 & \citet{Lunt1908} \\
1912.63 & 201.4 & 2.27 & \citet{Innes1914} \\
1926.02 & 200.6 & 2.74 & \citet{vandenBos1928} \\
1929.02 & 200.4 & 2.78 & \citet{Rossiter1955} \\
1936.74 & 201.2 & 2.75 & \citet{vandenBos1938} \\
1943.49 & 202.1 & 2.71 & \citet{Voute1955} \\
1956.446 & 200.83 & 2.831 & \citet{The1975} \\
1960.80 & 202.8 & 2.63 & \citet{vandenBos1961} \\
1966.81 & 201.0 & 2.82 & \citet{Knipe1969} \\
1975.722 & 201.0 & 2.79 & \citet{Worley1978} \\
1989.86 & 200.4 & 2.45 & \citet{Sahai1992} \\ 
2003.47 & 203.1 & 2.82 & HST (PI: Sahai) \\
\hline
\end{tabular}
\end{table}

\end{appendix}

\end{document}